\documentstyle[prl,aps]{revtex}
\input BoxedEPS.tex
\SetepsfEPSFSpecial
\HideDisplacementBoxes

\begin{document}

\twocolumn[\hsize\textwidth\columnwidth\hsize
          \csname @twocolumnfalse\endcsname
\title{Quasi-one-dimensional superconductivity above 300 K and quantum phase 
slips in individual carbon nanotubes} 
\author{Guo-meng Zhao$^{*}$} 

\address{ Department of Physics, Texas 
Center for Superconductivity~~and Advanced Materials, University of Houston, 
~\\
Houston, Texas 
77204, USA}

\maketitle
\widetext
\vspace{0.3cm}

\begin{abstract}
A great number of the existing data for electrical transport, the 
Altshuler Aronov Spivak  and Aharonov Bohm effects, as well 
as the tunneling spectra of individual carbon nanotubes  can be well explained by theories of the quantum phase slips in 
quasi-one-dimensional superconductors. The 
existing data consistently suggest that the mean-field superconducting 
transition temperature $T_{c0}$ in both single-walled and multi-walled carbon 
nanotubes could be higher than 600 K.  The quantum phase slip theories naturally explain 
why the on-tube resistances in the closely packed 
nanotube bundles or in the individual multi-walled nanotubes with large diameters 
approach zero  at room temperature, while a 
single tube with a small diameter has a substantial resistance.

~\\
~\\
\end{abstract}
\narrowtext
]
The discovery of high-temperature superconductivity 
at about 40 K in electron-doped C$_{60}$ \cite{Gunnarsson} 
suggests that the electron-lattice coupling with high-energy phonons in 
this graphite-related material should be strong. Otherwise, $T_{c}$ 
could not be so high because 
the Coulomb pseudopotential $\mu^{*}$ in C$_{60}$ should be larger 
than 0.2 due to a small Fermi energy ($\sim$ 0.2 eV) and a large phonon 
energy ($\sim$ 0.2 eV) \cite{Gunnarsson}. The strong electron-phonon coupling in C$_{60}$ 
may arise from the finite curvature of the graphite sheets, that leads 
to hybridization of  $\sigma$, $\pi$, $\sigma^{*}$ and $\pi^{*}$ 
states \cite{Blase} and thus enhances the electron-phonon coupling. 
Similarly, the finite curvature of the carbon nanotubes (CNTs) produces 
a stronger electron-phonon coupling compared to their  zero curvature 
counterpart, graphene.  Moreover, the CNTs have a quasi-one or 
quasi-two-dimensional electronic structure.  It was shown that in 
multi-layer systems such as cuprates and multi-walled nanotubes 
(MWNT), high-temperature superconductivity can occur due to an 
attraction of the carriers in the same conducting layer via exchange 
of virtual plasmons in neighboring layers \cite{Cui}.  Similarly, 
exchange of undampled acoustic plasmon modes in quasi-one-dimensional 
systems can also lead to high-temperature superconductivity 
\cite{Lee}.  Thus, the CNTs should have a much higher mean-field 
superconducting transition temperature $T_{c0}$ than doped C$_{60}$ 
due to the presence of strong electron-phonon and electron-plasmon 
coupling.  Zhao {\em et al.} \cite{Zhao} have recently argued for the 
existence of superconductivity above 600 K in MWNT ropes.  In order to 
confirm this claim, it is helpful to show the zero-resistance state at room 
temperature in the nanotubes.  However, one should also consider the 
quasi-one-dimensional nature of the nanotubes, which may lead to a 
finite on-tube resistance below $T_{c0}$.

Here we extensively analyze the existing data for electrical transport, magnetoresistance 
and tunneling spectra of both single-walled and multi-walled carbon 
nanotubes. The data can 
be well explained by theories of quantum phase slips (QPS)  in one-dimensional superconductors. 
Moreover, the existing data consistently suggest that the mean-field superconducting 
transition temperature $T_{c0}$ in both single-walled and multi-walled carbon 
nanotubes could be higher than 600 K.

It is known that superconducting fluctuations in one-dimensional 
(1D) superconductors play an essential role in the resistive 
transition. Slightly below superconducting transition temperature 
$T_{c0}$, 1D superconductors have a finite resistance due to thermally 
activated phase slips (TAPS) \cite{Langer}. 
A number of experiments have also demonstrated a large resistance 
well below $T_{c0}$ in thin superconducting
wires \cite{Giordano89,Giordano90,Giordano91,Tinkham}. Further, a crossover to an insulating state 
has  been observed in 
ultrathin PbIn wires with diameters of the order of 10 nm 
\cite{Giordano89,Giordano91} as well as in ultrathin wires of MoGe \cite{Tinkham}.

Theories based on the quantum phase slips can explain the finite resistance in 1D 
superconductors \cite{Giordano90,Zaikin}. Essentially, the phase slips at low temperatures are 
related to the macroscopic quantum tunneling (MQT), which allows  the  
phase of the superconducting order parameter to fluctuate between  zero 
and $2\pi$ at some points along the wire, resulting in voltage 
pulses. For a single electron, the scatterings by impurities, phonons and other electrons can 
change the phase of the electron. The 
uncertainty in the phase of the single electron due to scattering leads  to localization 
of the electron. By analogy, the uncertainty in the phase of the Cooper 
pairs due to QPS results in localization of the Cooper pairs and 
thus a non-zero resistance \cite{Tinkham}. The QPS tunneling rate is proportional to $\exp 
(-S_{QPS})$, where $S_{QPS}$ in clean superconductors is very close to the number of transverse 
channels $N_{ch}$ in the limit of weak 
damping (see below).  If the number of the transverse channels 
$N_{ch}$ is small, the QPS tunneling rate is not negligible, leading 
to a non-zero resistance at low temperatures.  For a single-walled 
nanotube (SWNT), $N_{ch}$ = 2, implying a large QPS tunneling rate and 
thus a large resistance even if it is a superconductor.  For MWNTs 
with several superconducting layers adjacent to each other, the number 
of the transverse channels will increase substantially, resulting in 
the large suppression of the QPS if the Josephson coupling among the layers 
is strong.  If two superconducting tubes are closely 
packed together to effectively increase the number of the channels, 
one would find a small on-tube resistance at room temperature if the 
constituent tubes have a mean-field $T_{c0}$ well above room 
temperature.  This can naturally explain why a single MWNT with a 
diameter $d$ of about 17 nm has a finite on-tube resistance at room 
temperature \cite{Sheo,Bachtold2000} while a bundle consisting of two 
MWNT tubes has a negligible on-tube resistance \cite{Frank}.

There are thermally activated 
phase slips (TAPS) and quantum phase slips in a thin superconducting wire. 
A theory developed by Langer, Ambegaokar, McCumber and Halperin 
\cite{Langer}, describes phase slips which occur via thermal activation.  The 
resistance due to the TAPS is given by \cite{Book}
\begin{equation}
R_{TA} = R_{Q}\frac{\hbar\Omega}{2k_{B}T}\exp (-\Delta 
F_{\circ}/k_{B}T),
\end{equation}
where $R_{Q}$ = $h/2e^{2}$ =12.9 k$\Omega$ is 
the resistance quantum and the attempt frequency $\Omega$ is given by 
\cite{Langer}
\begin{equation}
\Omega = \frac{\sqrt{3}}{2\pi^{3/2}}\frac{L}{\xi} \sqrt{\frac{\Delta 
F_{\circ}}{k_{B}T}}\frac{1}{\tau}.
\end{equation}
Here $L$ is the length of the wire, $\xi$ is the coherence length, and 
$\hbar/\tau = (8/\pi) k_{B}(T_{c0} -T)$. The barrier energy $\Delta F_{\circ}$ is 
\begin{equation}
\Delta F_{\circ} = \frac{8\sqrt{2}}{3}\frac{H_{c}^{2}}{8\pi}A\xi,
\end{equation}
where $H_{c}^{2}/8\pi$ is the condensation energy and $A$ is the 
cross-sectional area of the wire.  The condensation energy is equal to 
$N(0)\Delta^{2}/2$ within the BCS theory, where $N(0)$ is the average 
density of states near the Fermi level over the energy scale of the 
superconducting gap $\Delta$.  For a metallic SWNT with $N_{ch}$ =2, 
$N(0)A= 4/3\pi a_{C-C}\gamma_{\circ}$ (Ref.~\cite{RochePRB}), $\hbar 
v_{F} = 1.5a_{C-C}\gamma_{\circ}$ (Ref.~\cite{Mintmire}), where 
$\gamma_{\circ}$ is the hopping integral and $a_{C-C}$ is the bonding 
length.  Using $\xi = \hbar v_{F}/\pi\Delta$, $\hbar/\tau = (8/\pi) 
k_{B}T_{c0}$, $2\Delta/k_{B}T_{c0}$ =3.52, and the above relations, 
one can readily show that $\Delta F_{\circ}\tau/\hbar \simeq 0.13 
N_{ch}$ and $\Delta F_{\circ}/\Delta \simeq 0.19 N_{ch}$.  For MWNTs 
with $N_{l}$ superconducting layers, $\Delta F_{\circ}\tau/\hbar = 
0.26 N_{l}$ and $\Delta F_{\circ}/\Delta = 0.38 N_{l}$.

It was shown that the TAPS is significant only at temperatures very close 
to and below $T_{c0}$ \cite{Langer}. At lower temperatures, the finite resistance is caused by MQT and is given by 
\cite{Giordano90}. 
\begin{equation}\label{MQT1}
R_{MQT} =\beta_{1} R_{Q}\frac{L}{\xi}\sqrt{\frac{\beta_{2}\Delta 
F_{\circ}\tau}{\hbar}}\exp (-\beta_{2}\Delta 
F_{\circ}\tau/\hbar),
\end{equation}
where $\beta_{1}$ and $\beta_{2}$ are constants, depending on the damping 
strength. When the damping increases, $\beta_{2}$ decreases.

Substituting $\Delta F_{\circ}\tau/\hbar = 0.26N_{l}$ into 
Eq.~\ref{MQT1}, we find that
\begin{eqnarray}\label{MQT2}
R_{MQT} =\beta_{1}R_{Q}\frac{L}{\xi}\sqrt{0.26\beta_{2}N_{l} }~\nonumber \\
\exp (-0.26\beta_{2}N_{l}).
\end{eqnarray}

From Eq.~\ref{MQT2}, one can see that $S_{QPS} \simeq 2N_{l}$ in the 
limit of weak damping where $\beta_{2}$ = 7.2 (Ref.~\cite{Giordano90}). For a stronger 
damping, $\beta_{2}$ is reduced 
so that  $S_{QPS} < 2N_{l}$. Moreover, in the 
dirty limit, $S_{QPS}$ will be further reduced \cite{Zaikin} such that 
$S_{QPS} << 2N_{l}$.  
For a SWNT, $N_{l}$ = 1 so that a large QPS and a nonzero resistance is 
expected below the mean-field superconducting  transition 
temperature. If several 
superconducting SWNTs are 
closely packed to ensure an increase in the number of channels, the QPS would be 
substantially reduced. This can explain why the on-tube resistance is 
appreciable at room temperature for a single SWNT \cite{Soh,Yao} while the resistance at room 
temperature is very small for 
a bundle consisting of two strongly coupled SWNTs \cite{Bachtold2000}. 
For a MWNT with $d$ = 40 nm, there is a total of 27 metallic layers, that is, 
$N_{l}$ = 27 (Ref.~\cite{Pablo}). This implies that the QPS in this single MWNT should 
be strongly suppressed according to Eq.~\ref{MQT2}. Indeed, this MWNT 
has nearly zero on-tube resistance at room temperature over a length of 4 $\mu$m (Ref.~\cite{Pablo}).

More rigorous approach quantifying the QPS in quasi-1D superconductors 
\cite{Zaikin} suggests that $S_{QPS}$  depends not only  on the 
quantity $\Delta F_{\circ}\tau/\hbar$ but also on the normal-state 
conductivity $\sigma$ ($S_{QPS} \propto \sigma^{2/3}$). Therefore, 
one can very effectively suppress the QPS and the resistance below 
$T_{c0}$ by reducing 
the normal-state resistivity.  It was shown that the electron backscattering 
from a single impurity with long range potential
is nearly absent in metallic SWNTs  while this backscattering becomes significant 
for doped semiconducting SWNTs \cite{RocheAPL}.  This implies that the QPS in doped metallic 
SWNTs will be significantly smaller than that in doped semiconducting 
SWNTs  if both systems become superconducting by doping. 

Now we discuss the temperature dependence of the resistance observed 
in nanotubes. Fig.~1a shows  the temperature dependence of the four-probe resistance for a 
single MWNT with $d$ = 17 nm, which is reproduced from 
Ref~\cite{Sheo}. It is remarkable that the resistance increases with 
decreasing temperature, but saturates at low temperatures. This 
unusual temperature dependence is very difficult to be explained 
consistently in terms of the conventional theory of transport 
\cite{Sheo}. 
However, a theory based on the QPS in 1D superconductors can 
naturally explain this unusual behavior. It was shown that 
\cite{Zaikin},  the resistance $R \propto T^{3\mu-2}$ for $k_{B}T >> 
\Phi_{\circ}I/c$, and  $R$ becomes independent of temperature and is 
proportional to $I^{3\mu-2}$ for $k_{B}T << 
\Phi_{\circ}I/c$. Here $\Phi_{\circ}$ is the quantum flux, $c$ is 
the speed of light, $I$ is the current, and $\mu$ is a quantity that characterizes 
the ground state; the zero-temperature resistance can approach zero when $\mu > 2$, 
and is finite when $\mu < 2$.  The crossover from the power-law 
behavior  to the temperature independent 
behavior takes place at $T \sim \Phi_{\circ}I/ck_{B}$.  For 
example, the crossover temperature is about 7 K for $I$ = 50 nA.  When 
$\mu < 1.5$, $R$ increases with decreasing temperature (semiconducting 
behavior), while for $\mu > 1.5$, $R$ decreases with decreasing 
temperature (metallic behavior).  Only if the QPS are strongly 
suppressed, can a zero or negligible resistance state be realized below 
$T_{c0}$.

\begin{figure}[htb]
\input{epsf}
\epsfxsize 7cm
\centerline{\epsfbox{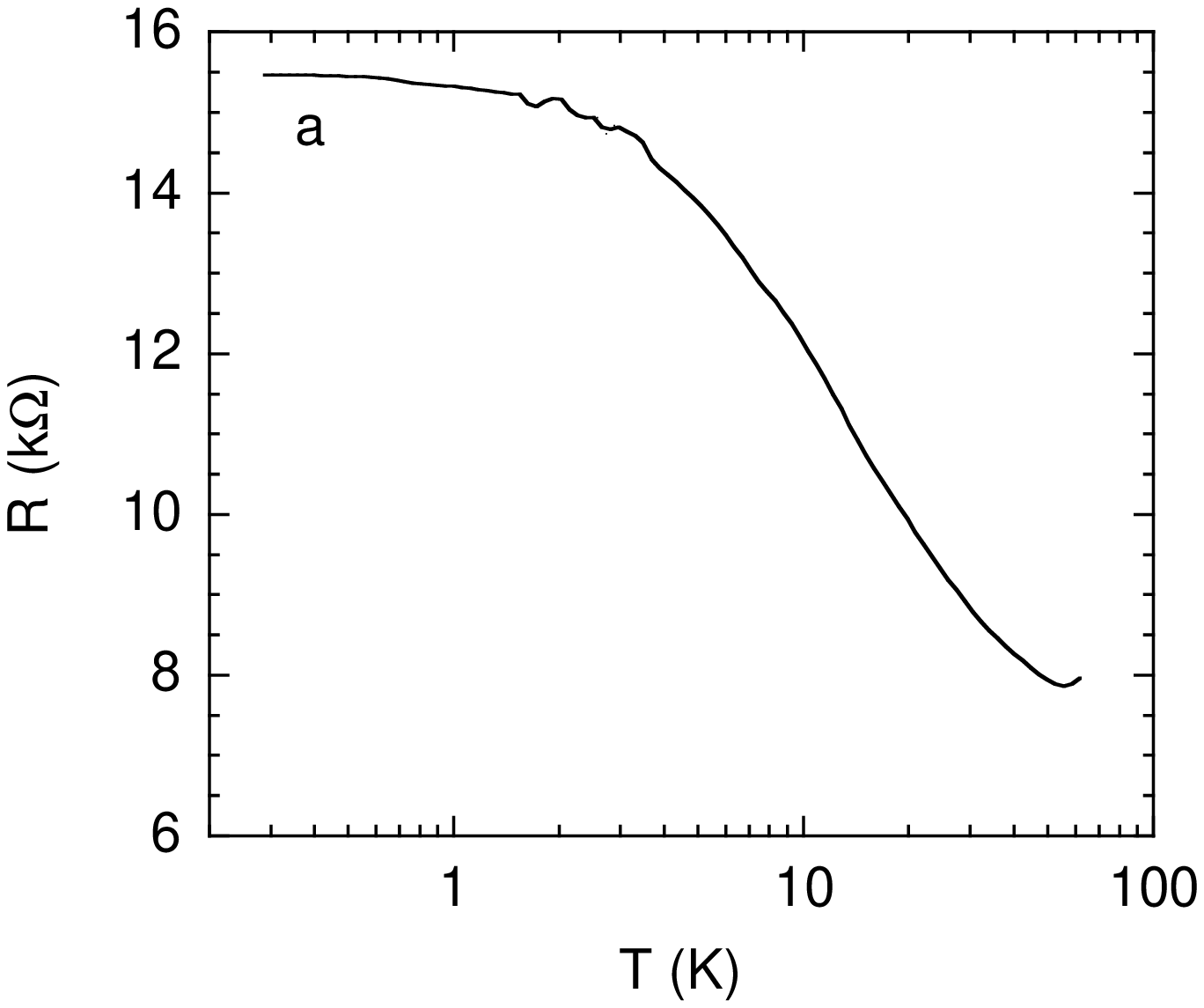}}
\vspace{0.4cm}
\input{epsf}
\epsfxsize 7cm
\centerline{\epsfbox{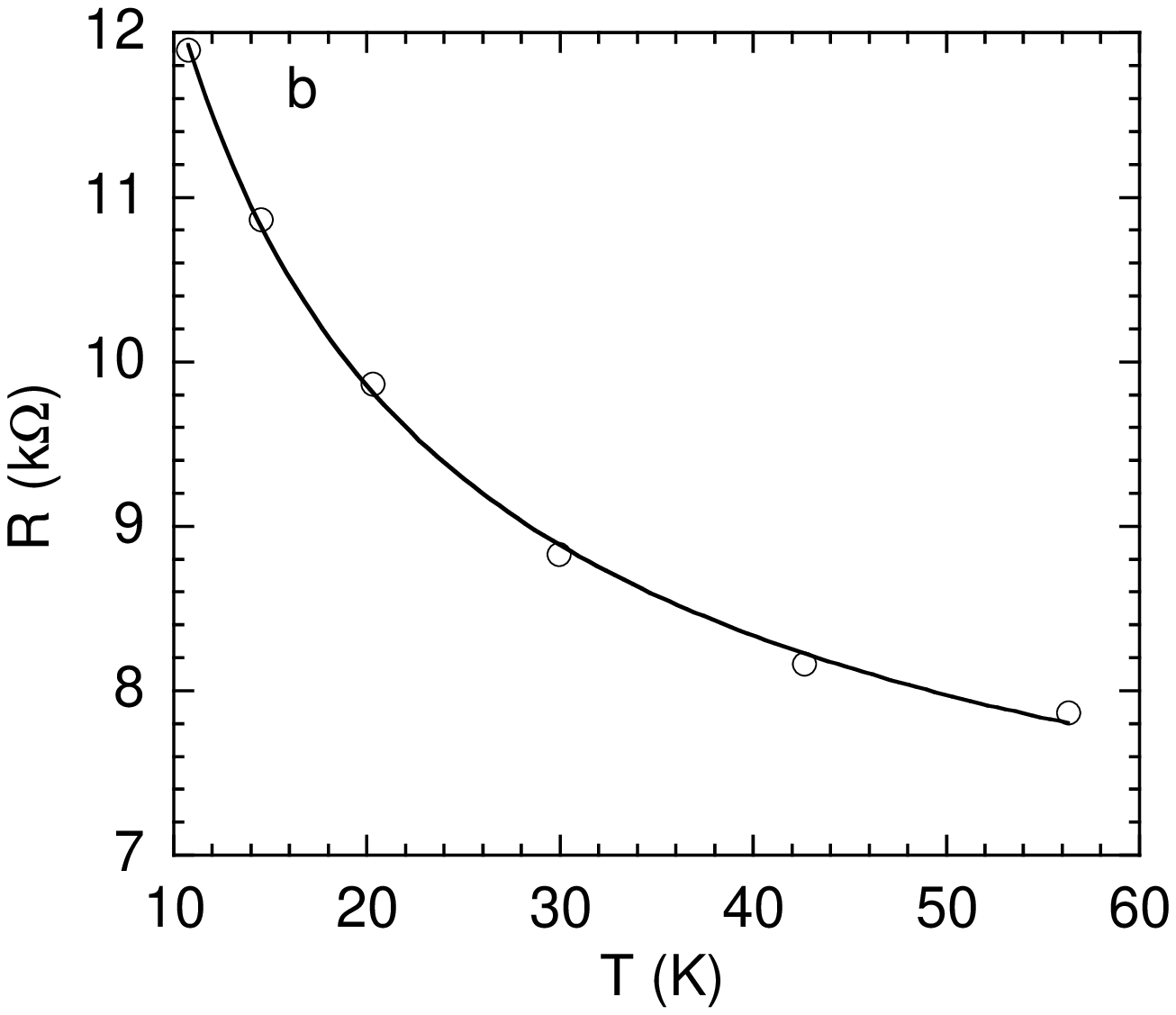}}
	\vspace{0.6cm}
	\caption [~]{a) The temperature dependence of the four-probe resistance for a 
single MWNT with $d$ = 17 nm, which is reproduced from 
Ref~\cite{Sheo}. b) The temperature dependence of the resistance over 
10-60 K, which is fitted by $R(T)  = R_{ct} + aT^{p}$.  The fitting 
parameters: $R_{ct} = 5.6(5)~k\Omega$, $p= -0.65(9)$, and $a = 
29(4)k\Omega K^{0.65}$. }
	\protect\label{fig1}
\end{figure}

From the QPS theory, we can show that for $\Phi_{\circ}I/ck_{B} << T << 
T_{c0}$, the four-probe resistance $R(T)$ is
\begin{equation}\label{R}
R(T)  = R_{ct} +aT^{p}, 
\end{equation}
where $R_{ct}$ is the tunneling resistance and $p = 3\mu-2$. The 
tunneling resistance is given by $R_{ct} = R_{Q}/tN_{ch}$, where $t$ is the transmission 
coefficient ($t \leq$ 1). 

In Fig.~1b, we fit the resistance of the MWNT by Eq.~\ref{R}.  The 
best fit gives $R_{ct}$ = 5.6(5)~k$\Omega$, $p= -0.65(9)$, and $a = 
29(4)~k\Omega K^{0.65}$. At zero temperature $R(0) = R_{s}+R_{ct}$, 
where $R_{s}$ is the on-tube saturation resistance at zero 
temperature. The value of $R_{ct}$ (5.6 k$\Omega$) and the value of 
$R(0)$ (15.3 k$\Omega$) suggest that the on-tube saturation resistance of the tube is 9.7 k$\Omega$. 
\begin{figure}[htb]
\input{epsf}
\epsfxsize 7cm
\centerline{\epsfbox{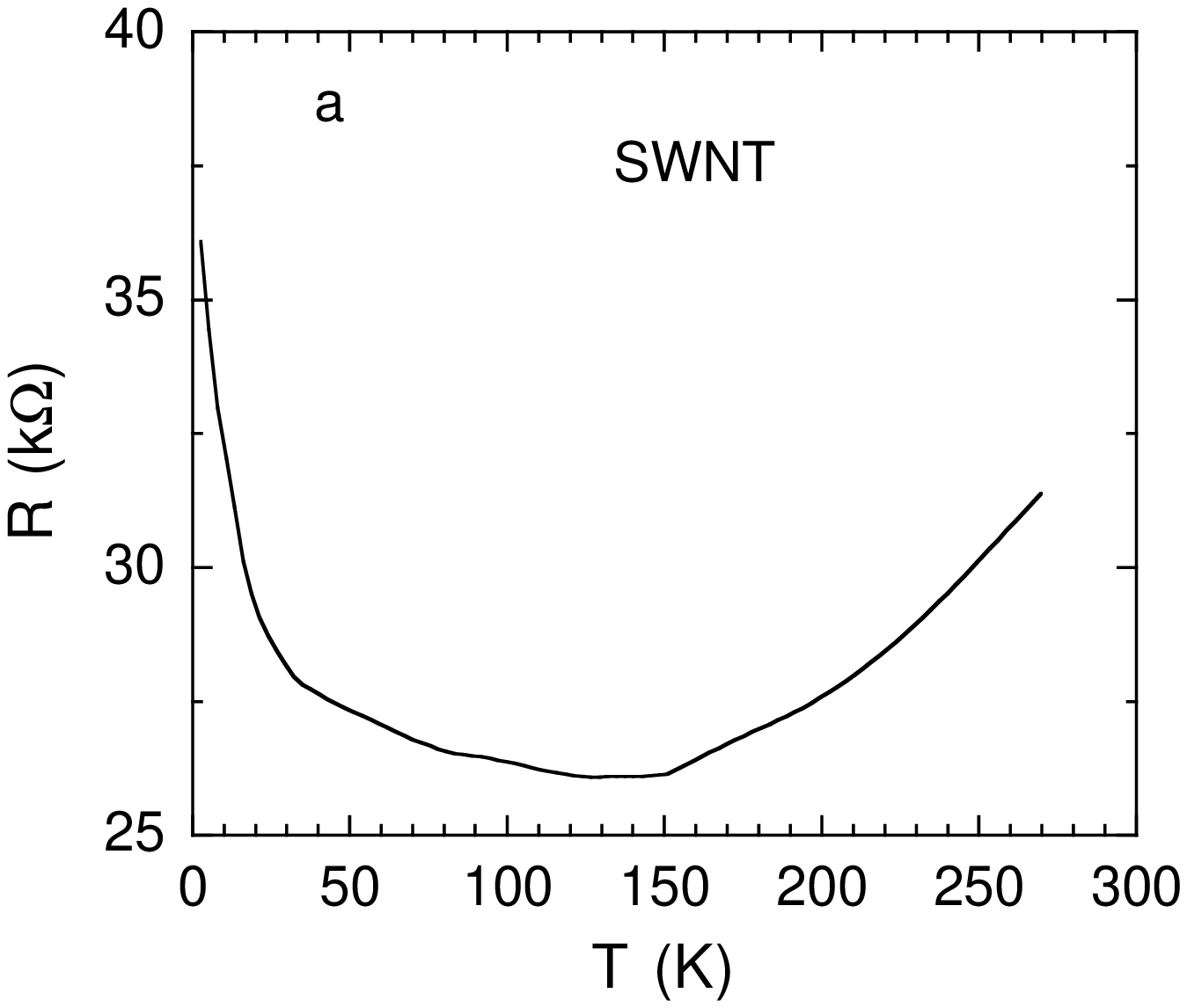}}
\vspace{0.4cm}
\input{epsf}
\epsfxsize 5.5cm
\centerline{\epsfbox{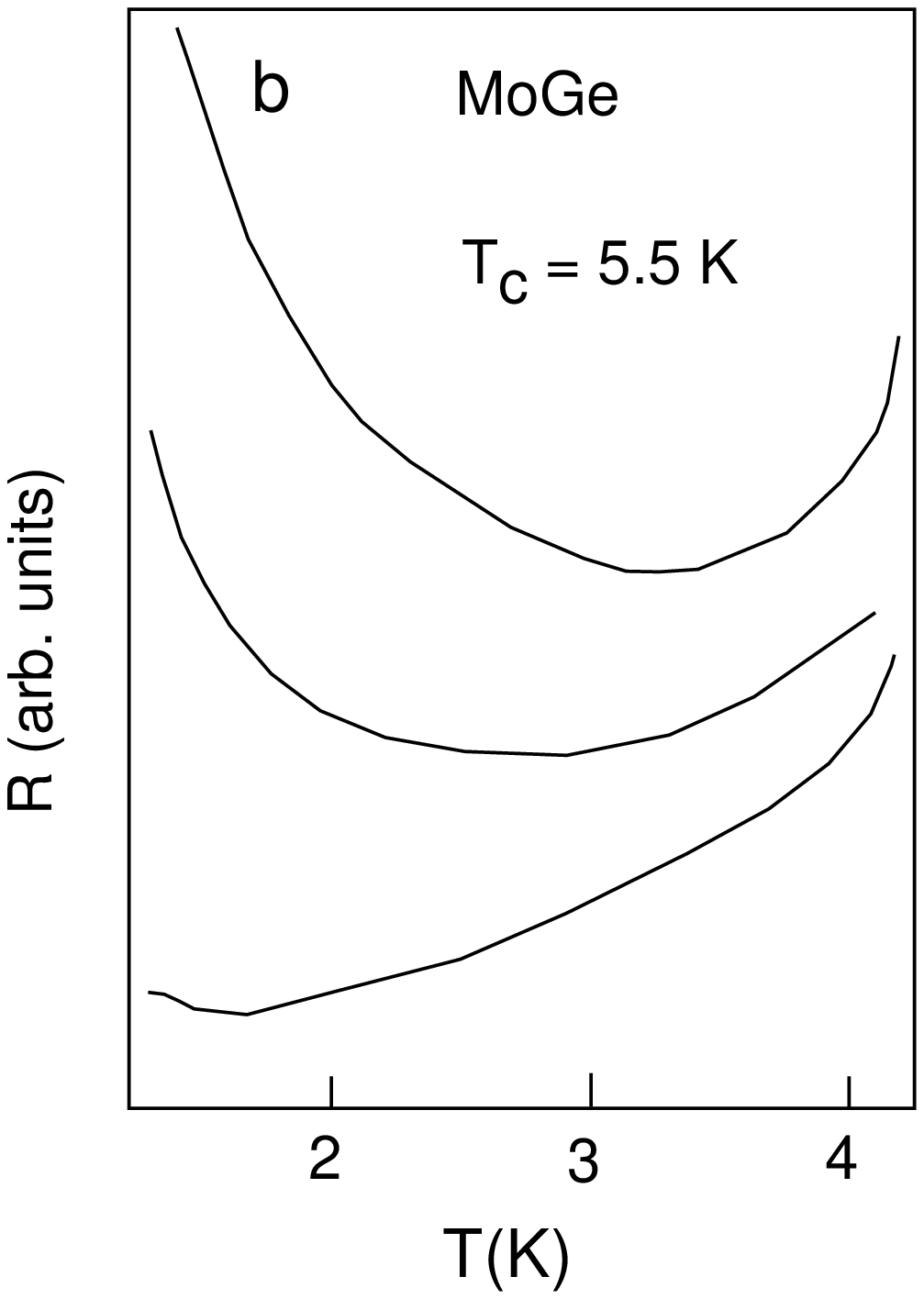}}
	\vspace{0.6cm}
	\caption [~]{a) Temperature dependence of the resistance for a 
SWNT. The data are extracted from Ref.\cite{Soh}.  b) Temperature 
dependence of the resistance for three ultrathin MoGe wires. The curves 
are smoothed from the original plot of Ref.~\cite{Tinkham}.}
\protect\label{fig2}
\end{figure}
From the value of $R_{ct}$ for the MWNT, 
we can estimate the transmission coefficient $t$.  As discussed below, the high bias transport 
measurements in MWNTs \cite{Collins} suggest that there is a total of 14 conducting 
layers in a MWNT with $d$ = 14 nm, and that the number of the 
conducting layers is nearly proportional to $d$. 
Then, the MWNT with $d$ = 17 nm should have about 17 conducting layers. 
Moreover, the beautiful experiment
reported by de Pablo {\em et al.} 
\cite{Pablo} indicates that each 
conducting layer contributes 1 transverse 
channel to electrical transport.  If one takes $N_{ch} =$ 17 for  the 
MWNT with d= 17 nm, one finds $t$ = 0.135.  This suggests that the tunneling 
is far from being ideal, which may arise from non-Ohmic contacts.  We 
should mention that each layer should contribute 2 channels if there 
were no interlayer coupling.  However, it has been shown that the 
interlayer coupling can significantly modify the electronic states 
near the Fermi level, leading to the modulation in the number of 
channels between 1 and 3 for each layer (the average number of 
channels over an energy scale of 0.1 eV remains 2 for each layer) 
\cite{Kwon,Sanvito}.

In Fig.~2a, we plot the temperature dependence of the resistance for a 
single SWNT. The data are extracted from Ref.\cite{Soh}.
It is interesting that the temperature dependence of the 
resistance in the single-walled nanotube is similar to that found for 
ultrathin wires of MoGe superconductors \cite{Tinkham}, facsimile of which is 
reproduced in Fig.~2b.  The characteristic 
temperature $T^{*}$ corresponding to the local resistance minimum depends 
on the resistance in the normal state.
It appears that $T^{*}$ decreases with decreasing resistance. The 
resistance at low temperatures could be smaller or larger than that in 
the normal state.  By comparing Fig.~2a and Fig.~2b, one might infer that 
the mean-field $T_{c0}$ of this nanotube is well above 270 K.

From the single-particle tunneling spectrum obtained through two 
high-resistance contacts (see Fig.~1b of Ref.~\cite{Yao}), we can clearly 
see a pseudo-gap feature which appears at an energy of about 220 
meV. The pseudo-gap feature should be related to the superconducting 
gap rather than to the Luttinger-liquid behavior, as shown below. Considering the broadening 
of the gap feature due to the large QPS and the double tunneling junctions in series, we estimate the 
superconducting gap $\Delta$ to be about 100 meV. Scanning tunnelling 
microscopy and spectroscopy \cite{Wildoer} on
individual single-walled nanotubes also show the pseudo-gap features 
with $\Delta \simeq$ 100 meV in doped metallic SWNTs (the Fermi level $E_{F}$ is about 
0.2 eV below the top of the valence band). Using $k_{B}T_{c0} = \Delta/1.76$, we 
find $T_{c0} \simeq$ 660 K. It is interesting to note that the pseudo-gap feature 
could also be explained by Luttinger-liquid theory \cite{Bock,Kane} 
assuming that the Luttinger parameter is far below the free fermionic 
value of 1.  The fact that the pseudo-gap feature is seen only in 
heavily doped metallic chirality tubes \cite{Wildoer} but not in 
undoped or lightly doped armchair tubes \cite{Lieber} may rule out the 
Luttinger-liquid explanation since theory \cite{Kane} does not expect 
that Luttinger-liquid behavior should disappear in undoped or lightly 
doped armchair tubes with finite lengths.

Now let us explain one of the most remarkable features observed in carbon nanotubes.  At large biases,  the current saturates at 19-23 
$\mu$A in SWNTs \cite{Yao}. The current saturation has been explained as 
due to the backscattering of the zone-boundary optical phonons  \cite{Yao}. 
However, the deduced mean free path for phonon backscattering is 
one order of magnitude smaller than the expected one from the 
tight-binding approximation \cite{Yao}. Further, the $I-V$ characteristic 
observed in SWNTs is temperature independent for $V >$ 0.1 V 
(Ref.~\cite{Yao}) while the calculated $I-V$ characteristic within 
this mechanism should strongly depend on temperature especially in the 
low-bias range \cite{Mahan}.

\begin{figure}[htb]
\input{epsf}
\epsfxsize 7cm
\centerline{\epsfbox{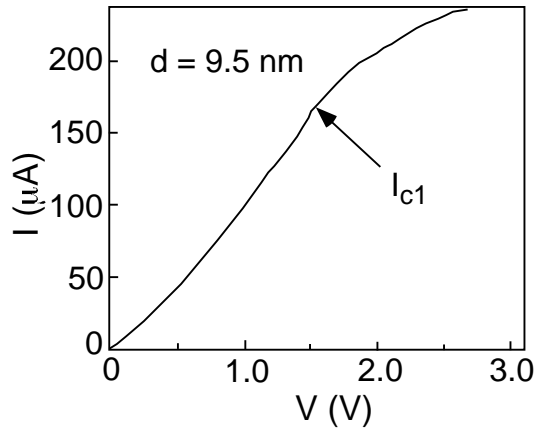}}
	\vspace{0.6cm}
	\caption [~]{The $I-V$ characteristic 
observed in a MWNTwith $d$ = 9.5 nm. The figure is reproduced from 
Ref.~\cite{Collins}.}
	\protect\label{fig3}
\end{figure}

 Alternatively, we explain the $I-V$ characteristics of both 
SWNTs and MWNTs \cite{Yao,Collins} in terms of quasi-1D 
superconductivity.  Essentially, the $I-V$ characteristics of both 
SWNTs and MWNTs \cite{Yao,Collins} are similar to that observed in 
ultrathin PbIn superconducting wires (see Fig~8 of 
Ref.~\cite{Giordano91}).  Fig.~3 shows the $I-V$ characteristic 
observed in a MWNT with $d$ = 9.5 nm.  The figure is reproduced from 
Ref.~\cite{Collins}.  When the applied current is below the lower 
critical current $I_{c1}$, the QPS are negligible so that the 
intrinsic on-tube resistance is much smaller than the normal-state 
resistance, and $V$ depends on $I$ rather linearly.  The slope $dV/dI$ is 
equal to the sum of the on-tube resistance and the contact resistance.  
When the applied current increases slightly above $I_{c1}$, the 
on-tube resistance rises rapidly towards the normal-state value due to 
the large QPS, leading to the dissipation that can burn the tube.  If 
the tube is not burned, the current tends to be saturated before the 
tube is completely driven into the normal state.  The saturation 
current is close to the mean-field critical current $I_{c}$ in the 
absence of defects.  Since phase slips initially occur near normal 
regions located around defects in the sample, we expect that $I_{c1}$ 
should strongly depend on the density of defects and thus on the 
normal-state resistivity.  From Fig.~3, one can see that $I_{c1}\leq 
0.75 I_{c}$.

According to the BCS theory, the mean-field critical current in the 
clean limit is given by \cite{Book}
\begin{equation}\label{Ic1}
I_{c}(T) = en_{s}(T)A\frac{\Delta (T)}{\hbar 
k_{F}}.
\end{equation}
The superfluid density $n_{s}(T)$ = $n 
\lambda^{2}(0)/\lambda^{2}(T)$, and the normal-state carrier 
density $n = 2N(0)E_{F}= 2N(0)\hbar v_{F}k_{F}$ = 
$4k_{F}/A\pi$.  Here we have used the relations: $N(0)A= 
4/3\pi a_{C-C}\gamma_{\circ}$, and $\hbar v_{F} = 
1.5a_{C-C}\gamma_{\circ}$, as well as $E _{F}= \hbar v_{F}|k_{F}|$.  
Substituting the above relations into Eq.~\ref{Ic1} yields
\begin{equation}\label{Ic}
I_{c}(T) = 
7.04N_{l}\frac{k_{B}T_{c0}}{eR_{Q}}\frac{\lambda^{2}(0)}{\lambda^{2}(T)}\frac{\Delta (T)}{\Delta(0)},
\end{equation}
with $I_{c}(0) = 7.04N_{l}k_{B}T_{c0}/eR_{Q}$.  Here 
$\lambda^{2}(0)/\lambda^{2}(T)$ follows the BCS prediction, and 
$\Delta (T) = \Delta (0)\tanh (1.6\sqrt{T_{c0}/T - 1})$, which is very 
close to that predicted by the BCS theory.  The critical current 
$i_{c}$ per superconducting layer is then given by
\begin{equation}\label{ic}
i_{c}(T) = 7.04\frac{k_{B}T_{c0}}{eR_{Q}}\frac{\lambda^{2}(0)}{\lambda^{2}(T)}\frac{\Delta(T)}{\Delta(0)}.
\end{equation}

\begin{figure}[htb]
\input{epsf}
\epsfxsize 7cm
\centerline{\epsfbox{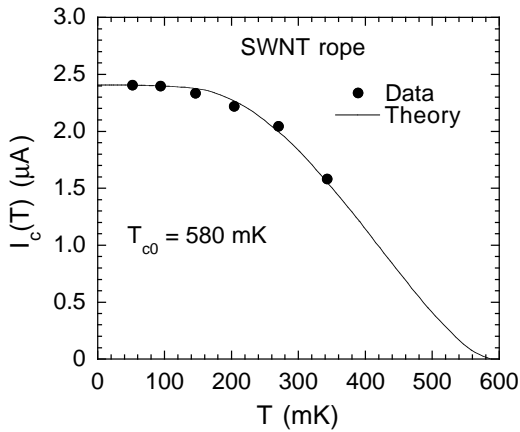}}
	\vspace{0.6cm}
\caption [~]{The critical current $I_{c}(T)$ for a SWNT rope.  The 
data are extracted from Ref.~\cite{Kociak}.  The solid line is the 
calculated curve using Eq.~\ref{Ic} and $T_{c0}=$ 580 mK.}
\end{figure}
For a SWNT rope, the resistance starts to drop below about 550 mK and 
reaches  a value $R_{r}$ = 74~$\Omega$ at low temperatures \cite{Kociak}.  The 
data are consistent with quasi-1D superconductivity with 
$T_{c0}\simeq$ 550 mK and $N_{l}$ = $R_{Q}/2R_{r}$ = 87 
(Ref.~\cite{Kociak}).  Substituting 
these numbers into the expression: $I_{c}(0) = 7.04N_{l}k_{B}T_{c0}/eR_{Q}$, we 
obtain $I_{c}(0)$ =2.25 $\mu$A, in excellent agreement with the measured 
$I_{c}(0)$ = 2.41~$\mu$A, as seen from Fig.~4.  The solid  
line in Fig.~4 is the calculated curve using Eq.~\ref{Ic} and 
$T_{c0}=$ 580 mK.  It is striking that the data are in quantitative 
agreement with theory.  We should mention that very low 
superconductivity in the SWNT rope may be due to the fact that the 
tubes are very lightly doped.  The very high normal-state resistance (830 
k$\Omega$/$\mu$m) per tube \cite{Kociak} suggests that the 
Fermi level must be very close to the top of the valence band where 
the Fermi velocity must be significantly reduced due to the opening 
of a small gap in non-armchair metallic tubes.

\begin{figure}[htb]
\input{epsf}
\epsfxsize 6.5cm
\centerline{\epsfbox{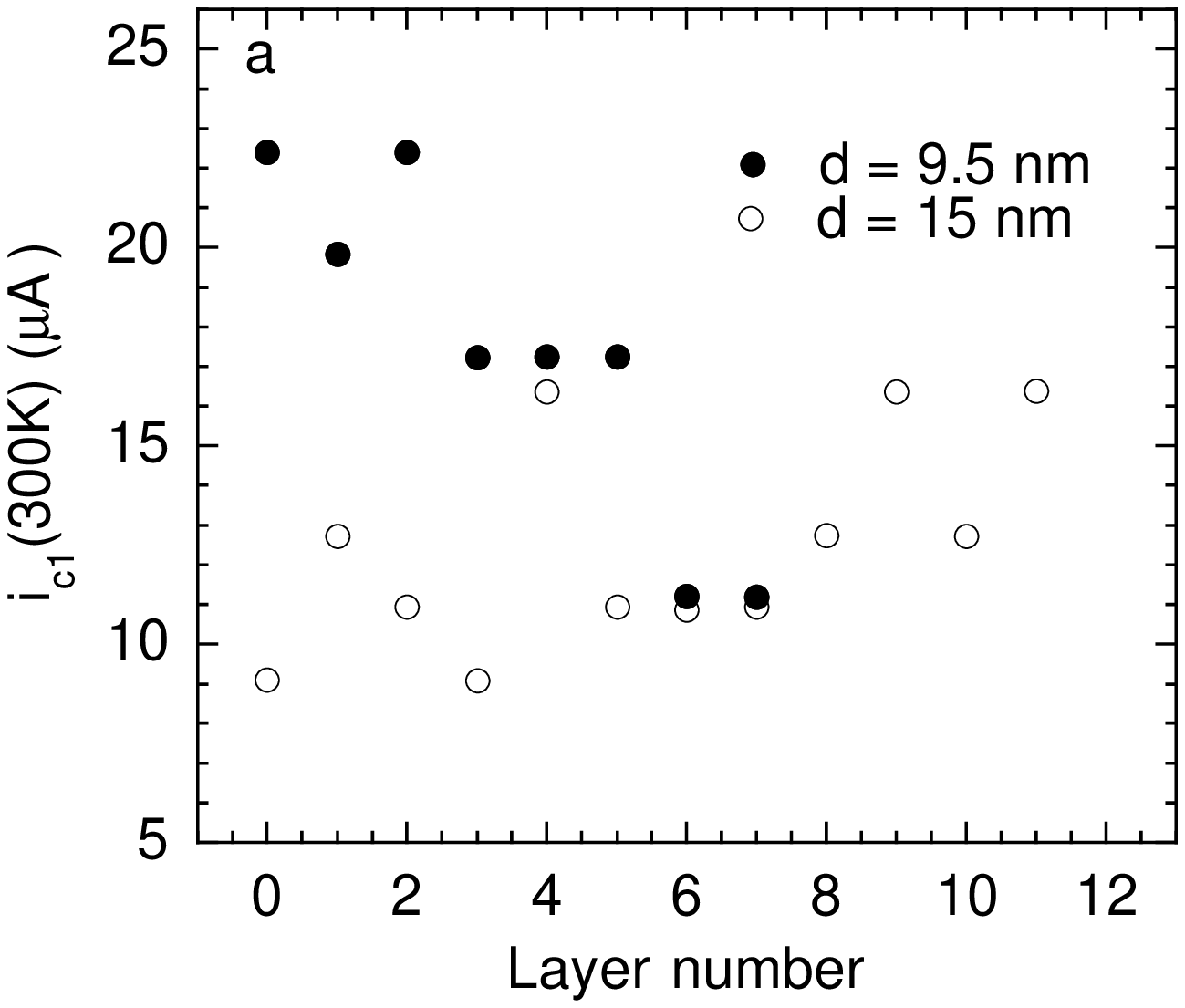}}
\input{epsf}
\epsfxsize 7cm
\centerline{\epsfbox{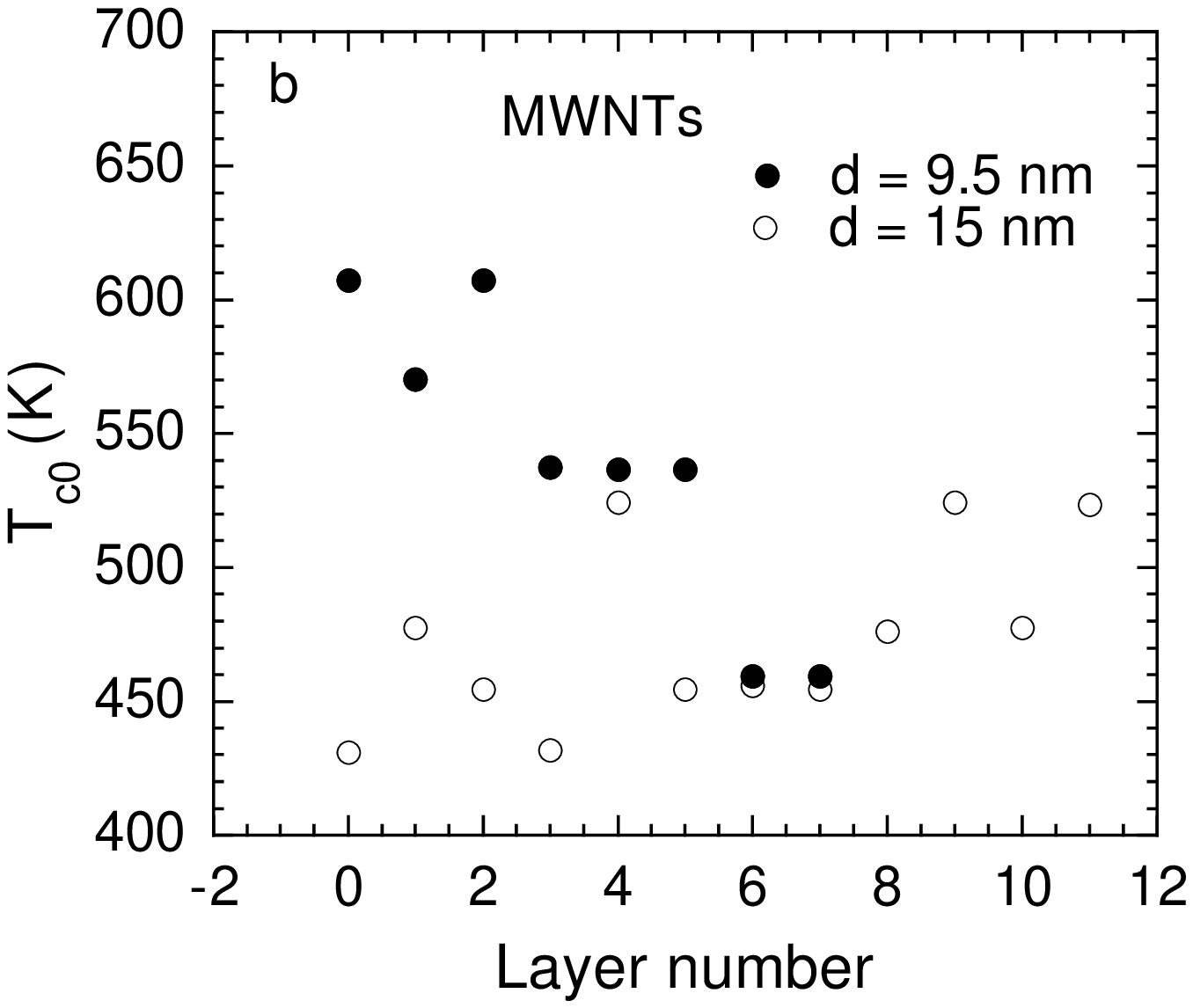}}
	\vspace{0.6cm}
	\caption [~]{a) The critical currents $i_{c}$'s at 300 K for the individual 
superconducting layers in a MWNT with $d$ = 9.5 nm and in a MWNT with $d$ = 
15 nm. The data are extracted from Ref.~\cite{Collins}. b) The mean-field critical temperature $T_{c0}$'s 
of individual superconducting layers in the MWNTs with $d$ = 9.5 nm 
and 15 nm, respectively. The $T_{c0}$'s  are
calculated using Eq.~\ref{ic} and assuming $i_{c} = i_{c1}$.}
\end{figure}

 In Fig.~5a, we show the critical currents $i_{c1}$'s at 300 
K for individual superconducting layers in a MWNT with $d$ = 9.5 nm 
and in a MWNT with $d$ = 15 nm.  The data are extracted from 
Ref.~\cite{Collins}.  The layer number starts from 0 that corresponds 
to the outermost superconducting layer.  For $d$ = 9.5 nm the $i_{c}$ 
tends to decrease with decreasing the diameter of the layer, while for 
$d$ = 15 nm the tendency is just the opposite.  Plotted in Fig.~5b is the 
mean-field critical temperature $T_{c0}$'s of individual 
superconducting layers in the MWNTs, which are calculated using 
Eq.~\ref{ic} and assuming $i_{c} = i_{c1}$.  From the discussion 
associated with Fig.~3, it is clear that the calculated $T_{c0}$'s are 
underestimated because $i_{c}$ should always be larger than $i_{c1}$.  One can see that $T_{c0}$ varies from 430 K to 610 
K, in good agreement with the independent resistance data \cite{Zhao}.  
This broad variation in $T_{c0}$ suggests that $T_{c0}$ depends on 
doping and on the diameters of tubes.  Indeed, a lower $T_{c0}$ value 
of about 300 K can be inferred from the temperature dependence of the 
resistance in a single MWNT with $d$ $\simeq$ 12 nm, which is 
reproduced from Ref.~\cite{Eb} and shown in Fig.~6.  It is likely that 
this MWNT is not optimally doped so that the $T_{c0}$ of the tube is 
much lower than the one ($>$ 600 K) for optimally doped MWNTs.

\begin{figure}[htb]
\input{epsf}
\epsfxsize 7cm
\centerline{\epsfbox{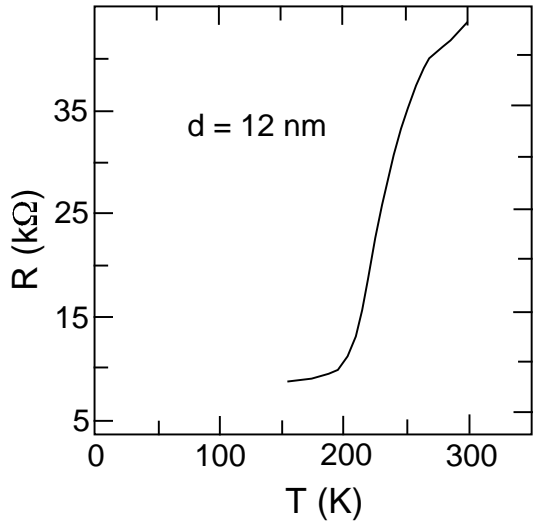}}
	\vspace{0.6cm}
\caption [~]{The 
temperature dependence of the resistance in a single MWNT 
with $d$ $\simeq$ 12 nm, which is reproduced from Ref.~\cite{Eb}. The 
sudden drop in the resistance below 300 K can be explained in terms of 
quasi-1D superconductivity with $T_{c0}$ $\simeq$ 300 K. The non-zero 
resistance below 200 K arises from phase slips and from the contact 
resistance $R_{ct}$ discussed above.}
\end{figure}

Now we turn to discuss the Aharonov Bohm (AB) effect, which has been observed in 
MWNTs when the magnetic field is applied along the tube-axis direction 
\cite{BachtoldNature,Sheo}. 
The magnetoresistance measurements
showed pronounced resistance oscillations as a function of magnetic
flux. The oscillation period was found to be about $\Phi_{\circ}$ (= 
$hc/2e$) if one assumed that only the outermost layer 
is involved in conduction \cite{BachtoldNature,Sheo}. The result could be consistent with  the Altshuler 
Aronov Spivak (AAS) effect, which arises from quantum interference of 
two counter-propagating closed diffusive electron
trajectories \cite{Aronov}. On the other hand,  a period of 2$\Phi_{\circ}$ should 
have been observed if the phase coherence length of single particles is 
reasonably larger than $\pi d$ (the AB effect for the single particle density 
of states \cite{Roche}).  If the phase coherence length $L_{\phi}$ deduced from 
experiment (e.g., $L_{\phi}$ $\sim$ 300 nm $>> \pi d$ in one of the MWNTs 
\cite{BachtoldNature}) were related to  that for single particles, one 
would have observed the AB effect. However, such an effect 
has never been observed \cite{BachtoldNature,Sheo}. Therefore, this contradiction cannot 
be resolved if the conduction carriers were single particles.

We can resolve the above discrepancy if we assume that the conduction 
carriers are Cooper pairs in the limit of weak localization (WL). As we 
discussed above,  the uncertainty in the phase of Cooper pairs due to 
the large QPS could lead to weak localization of the Cooper pairs 
\cite{Tinkham}.  In 
many situations, a Cooper pair can be equivalent to a particle with a
charge of 2$e$. This simplification suggests that the WL theory for single particles should 
be applicable for Cooper pairs upon replacing $e$ with $2e$.  With this 
interpretation, we can readily find that the magnetic-flux period of the AAS effect for the Cooper 
pairs is $\Phi_{\circ}/2$, and that the AB effect for the single 
particle density of states should be absent if the phase coherence length 
for single particles is less than $\pi d$. 

In fact, the assumption that only the outermost layer 
is conducting \cite{Sheo,BachtoldNature} is not justified. As we discussed above, 
14 and 27 layers are involved in conduction in MWNTs with $d$ = 14 nm 
and 40 nm, respectively.   Further, the resistance 
at 1.3 K for a MWNT with 
$d$ = 13 nm is 2.45 k$\Omega$ (Ref.~\cite{BachtoldNature}).  The value 
of the resistance suggests that there are at least 6 transverse 
channels and 6 conducting layers which are involved in conduction.  
The average magnetic flux sensed by the carriers in all the conducting layers 
should be $B\pi (r_{out}^{2}+r_{in}^{2})/2$, where $r_{out}$ and 
$r_{in}$ are the radii of the outermost and innermost conducting 
layers, respectively, and $B$ is the magnetic field.  We can calculate 
$r_{in}$ using the relation $r_{in} = r_{out} - 0.34 (N_{l}-1)$ nm.  
For the MWNT with $d$ = 17 nm, $N_{l}$ = 17 (see above), leading to 
$r_{in}$ = 3.06 nm.  From the measured magnetic-field period of 8.2 T 
in the MWNT \cite{Sheo}, we find that the magnetic-flux period is 
$0.51\Phi_{\circ}$, in quantitative agreement with the thesis that the 
charge carriers are Cooper pairs with a finite phase coherence length 
due to the QPS.

In summary, the theories based on the QPS in 1D superconductors can 
quantitatively explain a large number of the existing data for electrical transport, the AAS and AB 
effects, as well as tunneling spectra of both SWNTs and MWNTs. The 
existing data consistently suggest that the mean-field superconducting 
transition temperature $T_{c0}$ in both single-walled and multi-walled carbon 
nanotubes could be as high as 600 K.  It is 
interesting that the energy gap (pairing energy) in the carbon nanotubes is very close to that 
for deeply 
underdoped cuprates, which would exhibit phase-coherent superconductivity 
above room temperature if one could increase the superfluid density.  
The high pairing energy in both cuprates and carbon nanotubes may 
arise from strong electron-phonon coupling and strong electron-plasmon 
interaction.  The possibility of phase-coherent superconductivity 
above room temperature in a single MWNT with $d$ = 40 nm \cite{Pablo} 
is due to the fact that the effective mass for the graphite-related 
materials is more than two orders of magnitude smaller than in the 
cuprates, and that the QPS are strongly suppressed due to the large 
number of transverse channels.  ~\\
 ~\\

{\bf Acknowledgment:} I thank Dr. Pieder Beeli for his critical reading and 
comments on the manuscript. The author acknowledges financial support from the State of Texas 
through the Texas Center for Superconductivity and Advanced Materials 
at the University of Houston where some of the work was completed.  ~\\
~\\
* Correspondence should be addressed to gmzhao@uh.edu.

\end{document}